\documentclass[aip,rsi,reprint,graphicx]{revtex4-2} 

\usepackage{graphicx}
\usepackage{dcolumn}
\usepackage{bm}
\usepackage{textcomp}
\usepackage{amsmath,amsfonts,amssymb} 
\usepackage{subfigure}

\begin{document}


\title{Billion-pixel X-ray camera (BiPC-X)} 
\thanks{Contributed paper to the Proceedings of the 23rd Topical Conference on High-Temperature Plasma Diagnostics, Santa Fe, NM, USA, May 31 - June 4, 2020.
Rescheduled online, Dec. 14-17, 2020. Correspondence: (Z.W.) zwang@lanl.gov.}


\author{Zhehui Wang}
\affiliation{Los Alamos National Laboratory, Los Alamos, NM 87545, USA}
\author{Kaitlin Anagnost}
\affiliation{Dartmouth College, Hanover, NH 03755, USA}
\author{Cris W. Barnes}
\affiliation{Los Alamos National Laboratory, Los Alamos, NM 87545, USA}
\author{D. M. Dattelbaum}
\affiliation{Los Alamos National Laboratory, Los Alamos, NM 87545, USA}
\author{Eric R. Fossum}
\affiliation{Dartmouth College, Hanover, NH 03755, USA}
\author{Eldred Lee}
\affiliation{Dartmouth College, Hanover, NH 03755, USA}
\affiliation{Los Alamos National Laboratory, Los Alamos, NM 87545, USA}
\author{Jifeng Liu}
\affiliation{Dartmouth College, Hanover, NH 03755, USA}
\author{J. J. Ma}
\affiliation{Gigajot Technology, Pasadena, CA 91107, USA}
\author{W. Z. Meijer}
\author{Wanyi Nie}
\affiliation{Los Alamos National Laboratory, Los Alamos, NM 87545, USA}
\author{C. M. Sweeney}
\affiliation{Los Alamos National Laboratory, Los Alamos, NM 87545, USA}
\author{Audrey C. Therrien}
\affiliation{Universit\'e de Sherbrooke, Sherbrooke, QC J1K 2R1, Canada}
\author{Hsinhan Tsai}
\affiliation{Los Alamos National Laboratory, Los Alamos, NM 87545, USA}
\author{Xin Yue}
\affiliation{Dartmouth College, Hanover, NH 03755, USA}
%


\date{\today}

\begin{abstract}
The continuing improvement in quantum efficiency (above 90\% for single visible photons), reduction in noise (below 1 electron per pixel), and shrinking in pixel pitch (less than 1 micron) motivate billion-pixel X-ray cameras (BiPC-X) based on commercial CMOS imaging sensors.  We describe BiPC-X designs and prototype construction based on flexible tiling of commercial CMOS imaging sensors with millions of pixels. Device models are given for direct detection of low energy X-rays ($<$ 10 keV) and indirect detection of higher energies using scintillators. Modified Birks's law is proposed for light-yield nonproportionality in scintillators as a function of X-ray energy. Single X-ray sensitivity and spatial resolution have been validated experimentally using laboratory X-ray source and the Argonne Advanced Photon Source. Possible applications include wide field-of-view (FOV) or large X-ray aperture measurements in high-temperature plasmas, the state-of-the-art synchrotron, X-ray Free Electron Laser (XFEL), and pulsed power facilities. 
\end{abstract}


\maketitle 

\section{Introduction}\label{intro}
\begin{figure}[htbp] 
  \centering
   \includegraphics[width=3.0in]{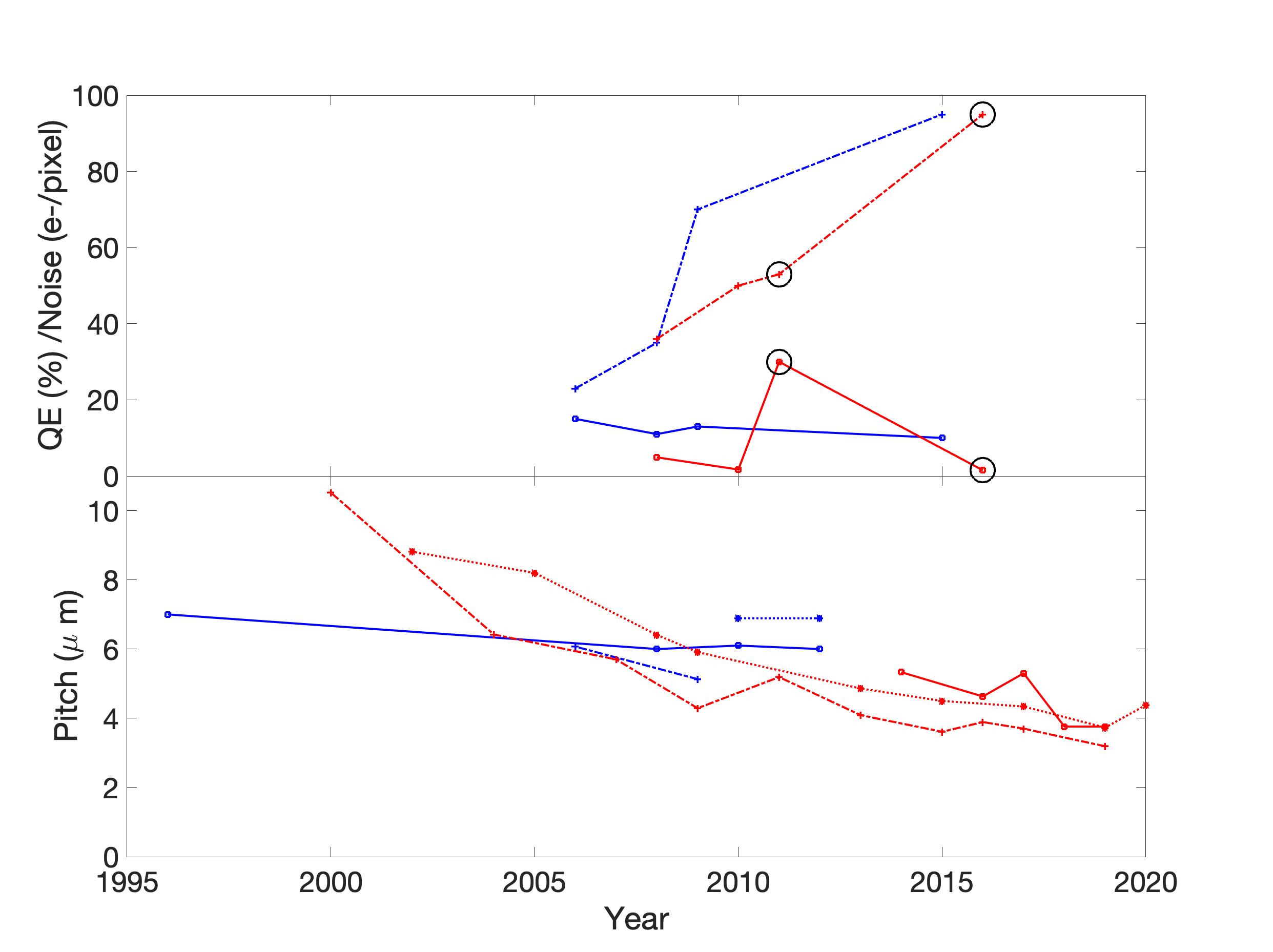} 
   \caption{A brief survey of the evolutionary trends of CCD (in blue) and CIS (in red) over the last 25 years. The quantum efficiency (QE) for visible photons has now exceeded 90\%. The noise level per pixel continues to decline, reaching 1 electron per pixel per readout cycle or less. Individual pixel size or pitch is  $<$ 5 $\mu$m as of 2020. These performance trends, in combination with continuing decline in cost, allow flexibility in BiPC-X camera designs and applications.}
  \label{fig:CMOS1}
\end{figure}

Room-temperature Complementary Metal Oxide Semiconductor (CMOS) imaging sensors have entered the single-visible-photon-sensitive regime without avalanche gain~\cite{MMS:2017}. Uses in personal devices such as cell phones and growing applications in machine vision have continuously pushed performance improvements, Fig.~\ref{fig:CMOS1}, and cost reduction for CMOS imaging sensors (CIS). As a result, CIS have gradually taken over charge coupled devices (CCD) imaging sensors over the last decade. Compared with CCD, which are serial devices when light-induced charge is read out one pixel at a time, row/column by row/column, CIS are based on parallel pixel architecture, when all pixels are designed to be exactly the same, including the readout electronics. Since electric charge from each pixel can be read out in parallel, CIS are better suited for high-speed applications than CCD. Consumer CIS have already reached 1000 frames per second (fps). One of the main results here is that high-performance low-cost visible-light CIS open door to billion-pixel X-ray camera (BiPC-X) designs, which may find applications such as in wide field-of-view measurements of high-temperature plasmas, pulsed power facilities, and X-ray scattering experiments in the state-of-the-art light sources including synchrotrons and X-ray free electron lasers. There are several approaches to overcome the low detection efficiency of the visible-light CIS for X-ray photon detection. A multi-layer CIS architecture has been described recently~\cite{Wang:2015,Drag:2016}, and validated with initial X-ray experiments at the Argonne Advanced Photon Source (APS)~\cite{Li:2019}.  Another approach is to integrate photon energy attenuation layers (PALs) with CMOS at pixel level~\cite{Lee:2009}. Alternatively, we may enhance the X-ray efficiency of each CIS by a scintillator converter. The latter approaches can also be extended to a multilayer configuration.

X-ray Bremsstrahlung and characteristic line emissions from impurity ions are signatures of keV and higher temperature plasmas. Recent advances in data-driven science offer new toolboxes such as neural networks to diagnose and understand high-temperature plasmas through three-dimensional (3D) X-ray imaging and tomography. Diffusive X-ray emissions from plasmas and the need to capture a large amount of X-rays data for applications such as training of deep neural works motivate BiPC-X or a giga-pixel X-ray camera instrument. One of the first giga-pixel cameras, AWARE-2, was reported in 2012 for visible light imaging~\cite{BGS:2012}.  AWARE-2 used a 16-mm entrance aperture to capture one-gigapixel images at three frames per minute. The Large Synoptic Survey Telescope (LSST) camera has 3.2 billion pixels by tiling 189 CCDs and a 0.5-fps frame rate. A growing number of billion-pixel visible cameras has since been reported.

Here we describe the design studies and initial results towards a BiPC-X. Sec.~\ref{sec:TP} is on the designs based on tiling of commercial CIS with millions of pixels and prototype construction using 3D printing of multi-sensor frame. In Sec.~\ref{sec:DM}, device models are given for direct and indirect detection of X-ray photons. It is found that above 10\% efficiency can in principle be obtained using the CMOS photo-diodes directly for photon energies below 10 keV. Modified Birks's law is proposed for scintillator light yield. Sec.~\ref{sec:ex} summarizes the experimental results on sensitivity and resolution. Follow-on work includes application in plasmas and further optimization of BiPC-X prototype design and performance.
\section{Design \& prototype}\label{sec:TP}
\begin{figure}[htbp] 
  \centering
   \subfigure[]{
   \includegraphics[width=2.8in]{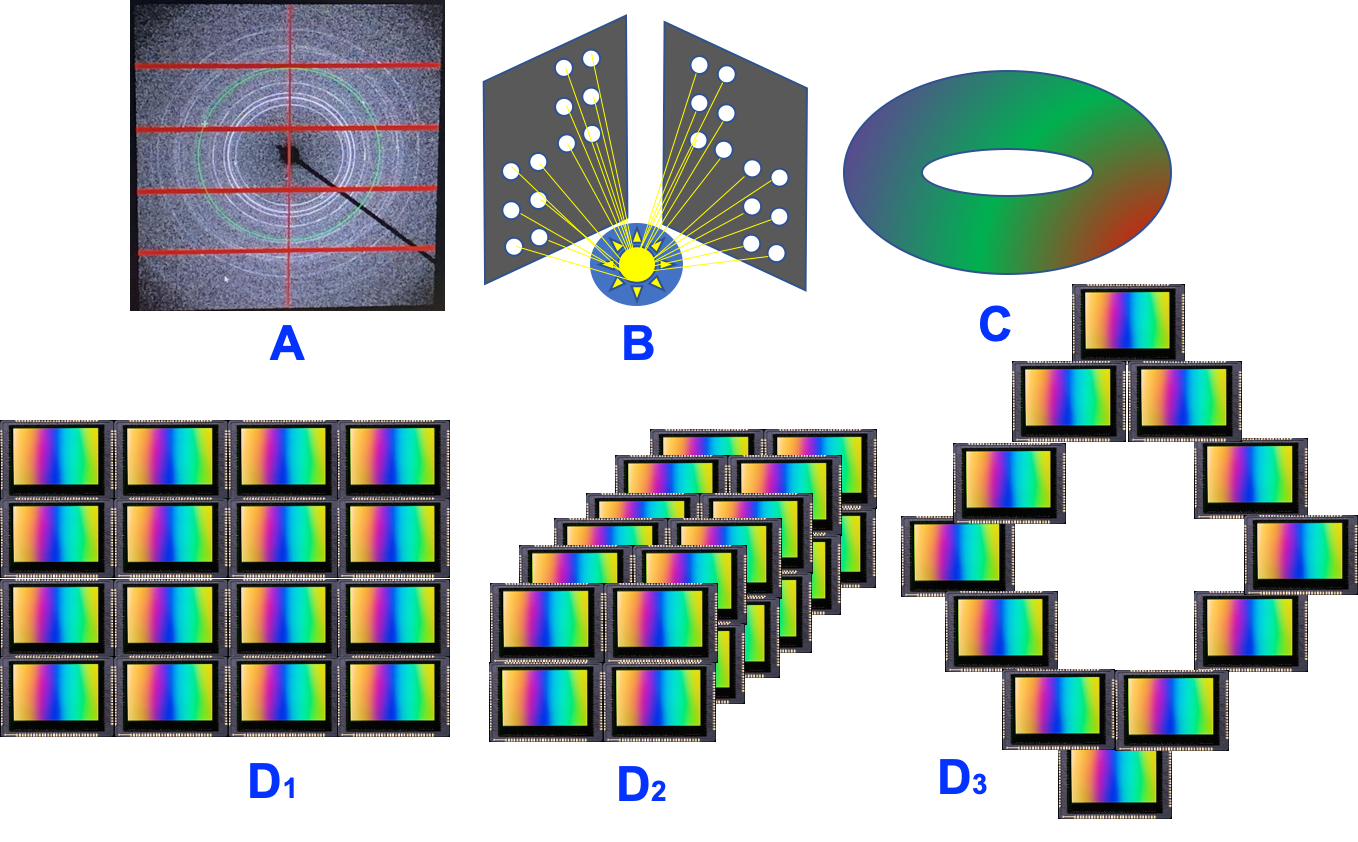}}
   \subfigure[]{\includegraphics[width=2.2in]{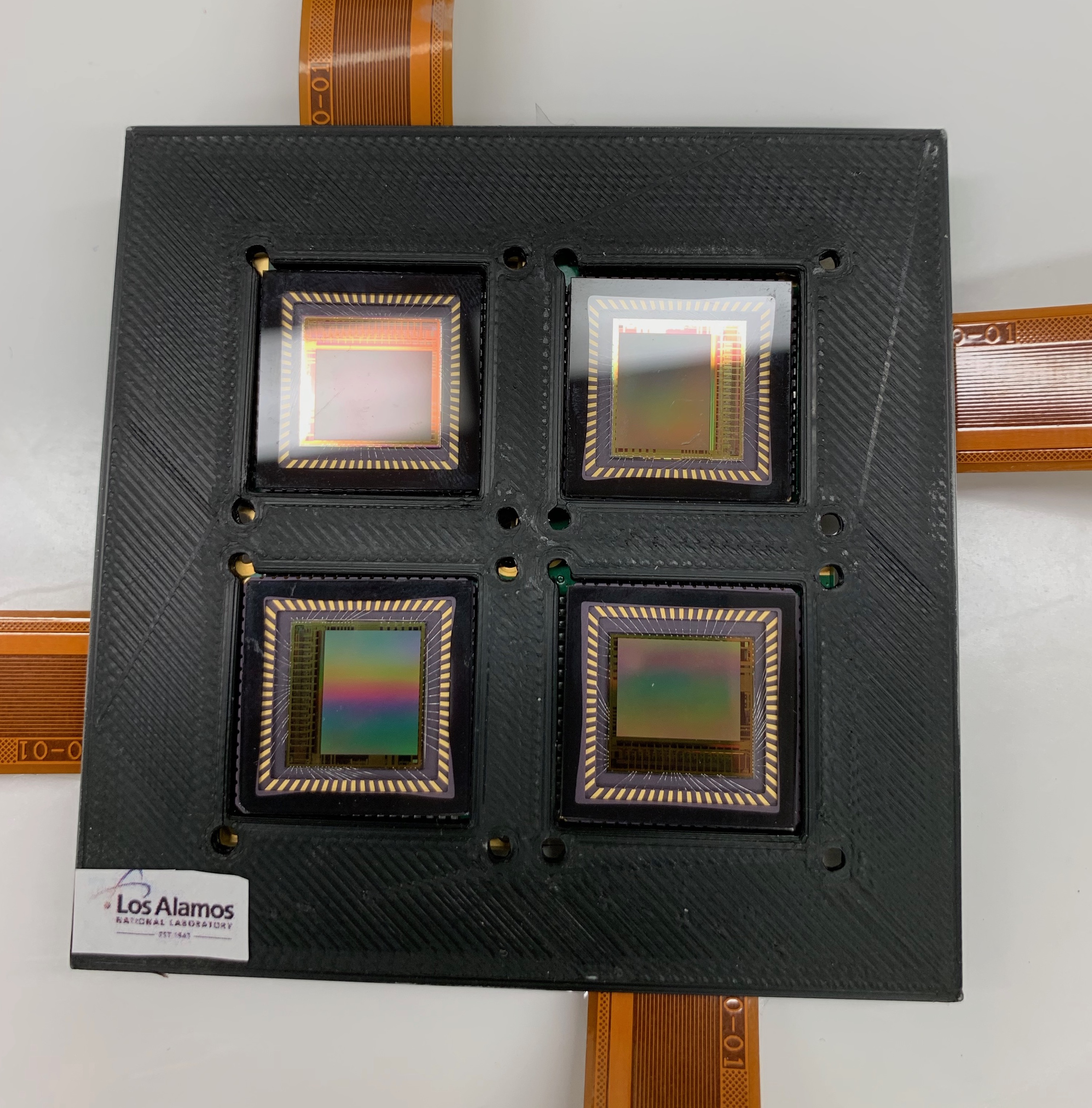}
  }
       \caption{(a) A BiPC-X may find applications in X-ray diffraction (A), inertial confinement fusion (B) and magnetic fusion (C). Examples of stacking and tiling to form a BiPC-X: Planar compact tiling configuration (D$_1$), stacked tiling (D$_2$),  and distributed tiling (D$_3$). (b) A laboratory 2$\times$2 tiling prototype using four On Semi Vita 5000 CIS.}
  \label{fig:App1}
\end{figure}

Using as building blocks the CIS with millions of pixels (MP), a BiPC-X can be constructed through multi-layer stacking and tiling~\cite{Wang:2015,Drag:2016,Li:2019}. Several possible configurations are illustrated as D$_1$, D$_2$ and D$_3$ in Fig.~\ref{fig:App1}. The planar compact tiling configuration D$_1$ increases the X-ray detection aperture, which is proportional to the number of CIS and the individual sensor area. The stacked tiling configuration D$_2$ increases the aperture for high-energy X-rays above 20 keV that can penetrate through multiple layers of CIS. High-energy X-rays and gamma rays ($\sim$ MeV) are expected from run-away electrons in tokamaks and by nuclear fusion. Configuration D$_3$ can be used in a toroidal plasma device such as a tokamak or a stellarator.  The synchrotron radiation from run-away electrons in a torus, as well as the bulk X-ray emissions can be captured by the CMOS sensor arrays surrounding the plasma in the poloidal plane. 

There are a large number of commercial CIS to choose from, and they differ in the total number of pixels, pixel pitch, speed, and cost. The latest models offer 10s of MP. Examples include Samsung's ISOCELL Bright HMX sensor (108 MP), the Canon 120MXS (122 MP), Gpixel’s GMAX3005 (150 MP),  OmniVision's OV64C (64 MP), and  ON Semiconductor's XGS 45000 (44.7 MP). A 5$\times$5 array of such sensors would be sufficient for a BiPC-X, with a pixel resolution below 1 $\mu$m except for GMAX3005 (5.5 $\mu$m, rolling shutter) and XGS 45000 (3.2 $\mu$m, global shutter). The frame rate of such a BiPC-X would be limited to about 1k fps for now, depending on the CIS. For pinhole imaging and tomography of inertial fusion plasmas, the kfps frame rate of such a camera can be compensated by a.) Using the gated scintillator and micro-channel plate (MCP) frontend; or b.) Exposure time gating of the CIS. In both cases, one or several cameras would capture one fast (1 $\mu$s or shorter exposure time) X-ray image. A fast X-ray movie would be generated by gating the sensors with different pre-programmed time delays. Additional customization of the CIS may be possible by increasing the X-ray sensitive region. The direct X-ray detection efficiency of the commercial off-the-shelf CIS is below 10\%, limited by the pinned photodiode dimension in each pixel to 2 - 5 $\mu$m (the photodiode depth should be larger than 3 $\mu$m to ensure red sensitivity~\cite{TWU:2014}) and the small CMOS operating bias voltage of several volts~\cite{FH:2014}. There are rooms to substantially increase the photodiode depth to hundreds of microns, making such a photodiode efficient for X-ray energies up to 10 keV, and thus suffice for many laboratory high-temperature plasmas. CIS are currently manufactured on 200 mm to 300 mm Si wafers. A standard 200 mm silicon wafer has a thickness of 725 $\mu$m. A 300 mm silicon wafer has a thickness of 775 $\mu$m. Current visible light CIS only use a small fraction of the wafer thickness, less than 10 $\mu$m.

A laboratory 2$\times$2 tiling prototype (21 MP total) using four ON Semi Vita 5000 CIS (5.3 MP, 75 fps, global shutter, 4.8 $\mu$m pitch, mono, die thickness 750 $\mu$m, glass lid thickness 550 $\pm$ 50 $\mu$m) has been built, Fig.~\ref{fig:App1}b. We used a 3D printer (Lulzbot Taz 6) to make the mounting frame for the 4 CIS.  The Fused Filament Fabrication printing method used  PolyMax PLA filament (from Polymaker). The thickness of the frame printed is 0.100$''$ to allow the detector to slightly protrude beyond the frame. Although the base circuit board is a 1.27$''$ square, the imaging detector is slightly rectangular and offset from the center of the chip. This requires consideration of how detectors will be oriented (for any size array) to ensure there is adequate room for the attached circuitry. As they are now, the imaging detectors are required to be at least 0.32$''$ apart to allow room for the boards they are attached to without overlapping with one another. Using the Lulzbot Taz 6 printer, a monolithic frame for up to 8$\times$8 (339 MP, Vita 5000) can be printed at once within a few hours. Frames for a BiPC-X are feasible with a larger printer or using a sensor with 16 MP, such as VITA16K from ON Semiconductor. 
\section{Device models \label{sec:DM}}
\begin{figure}[htbp] 
  \centering
   \includegraphics[width=3.0in, angle=0]{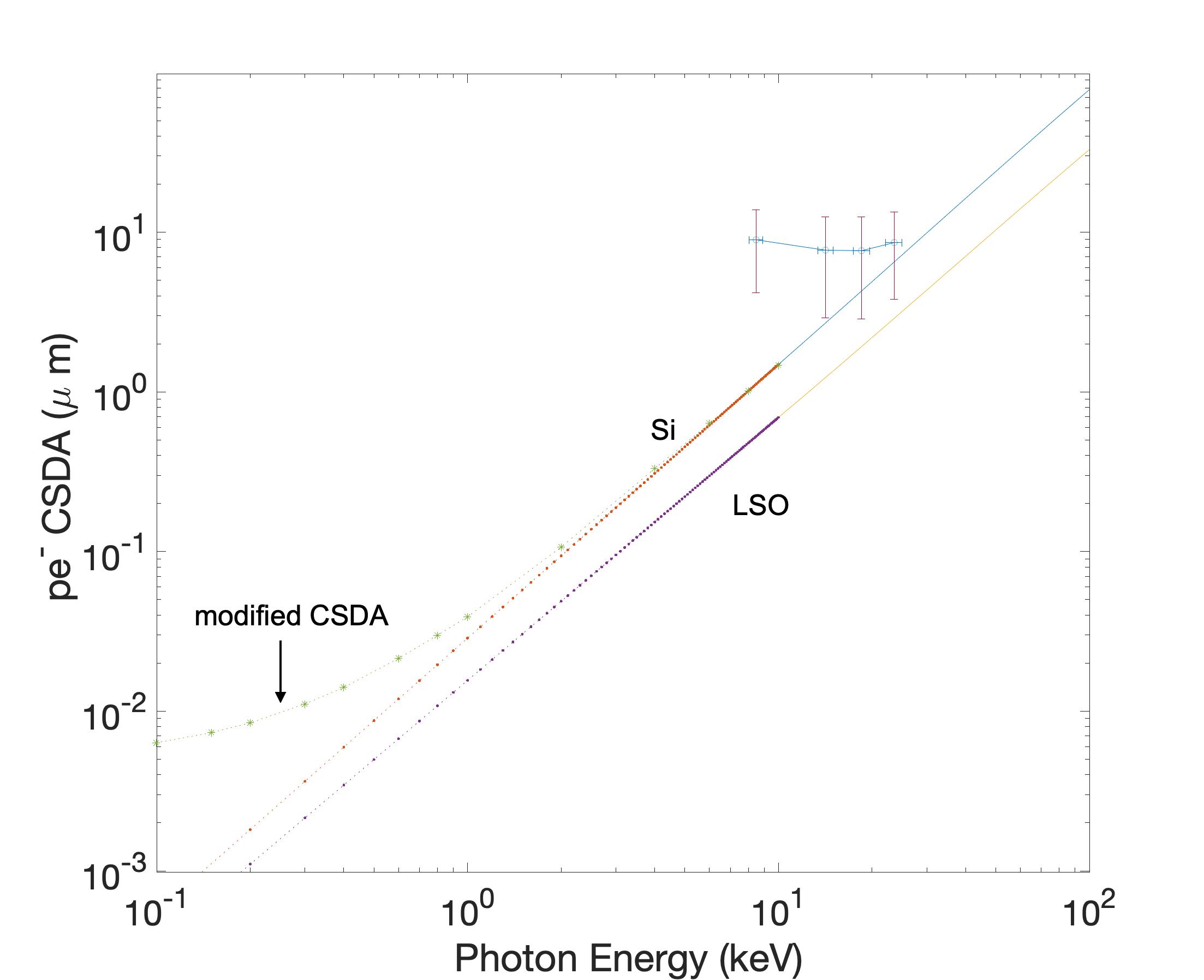} 
   \caption{Photo-electron range in Si and LSO scintillator as a function of X-ray energy based on CSDA model. Modification to CSDA model for Si is also included for $E_X < 10$ keV. Experimental data using an variable energy X-ray source indicates that resolution for direct single X-ray photon detection is mainly determined by charge sharing among neighboring pixels. The horizontal error bar corresponds to the energy spread of the X-ray source. The vertical error bar corresponds to 1 pixel width of 4.8 $\mu$m of the Vita 5000 CIS.}
  \label{fig:csda1}
\end{figure}

Here we describe device models for single X-ray photon detection efficiency and sensitivity. The response time is currently limited at CIS. The analysis provides theoretical basis for BiPC-X component selection and understanding of the component testing data described in Fig.~\ref{fig:csda1} and in Sec.~\ref{sec:ex}, especially the CIS and scintillators. Overall system performance parameters such as detective quantum efficiency (DQE), resolution or blur characterized by modulation transfer function (MTF) may also be derived, which is not included below partially due to the observation that the X-ray source properties, X-ray source, object and detector standoff distances could also play a role and thus need additional setup information~\cite{WWH:2017}. 

Device models may be divided into direct detection schemes based on X-ray attenuation in silicon photodiodes in CIS and indirect detection scheme with the primary X-ray attenuators being scintillators. The direct detection is more suitable for X-ray energies up to about 10 keV. The 1/$e$ attenuation length in silicon is 2.7, 17.5, 127, and 962 $\mu$m for 1, 5, 10, and 20 keV. Correspondingly, the fraction of X-ray attenuation and therefore the detection efficiency decreases from 82.9\%, 24.8\%, 3.9\% to 0.5\% in a silicon pinned photodiode of thickness 5 $\mu$m. At 20 keV, the $1/e$ attenuation length in silicon exceeds the 300 mm silicon wafer thickness of 775 $\mu$m. We shall mention without elaboration that other materials and structures typically used in CIS such as the glass lid have non-negligible effects on X-ray detection efficiency for energies below 20 keV.

We consider the planar compact tiling configuration, D$_1$ in Fig.~\ref{fig:App1}, which is sufficient for X-ray energies below 20 keV and plasmas with comparable or lower temperatures. The direct detection model in silicon involves X-ray to electron conversion, electron-hole (e-h) cloud propagation, and noise model for the device. In silicon photodiodes, 20 keV X-ray photoelectric (PE) absorption (91.6\%) dominates over other processes such as Compton scattering (3.1\%) and coherent scattering (5.3\%). The  PE fraction is more than 97\% for X-ray energies less than 10 keV. Based on the continuous slowing-down approximation (CSDA) and its modification at lower energies ($<10$ keV),  Fig.~\ref{fig:csda1}, the initial charge (e-h pairs) cloud produced from the energetic electrons ($\leq$ 20 keV) generated from PE process does not exceed 4.9 $\mu$m, which is comparable to the Vita 5000 CIS pitch of 4.8 $\mu$m. The number of e-h pair created can be estimated as $N_{eh} = E_X / E_0 \pm  \sqrt{f_0 E_X/E_0}$ for X-ray energy $E_X$. $E_0$ is 3.64 eV, and Fano factor $f_0$ is 0.13 for silicon. At $E_X=5$ keV for example, $N_{eh} = 1374 \pm 13$. Further spread of the charge cloud is due to e-h diffusion in silicon and charge sharing among multiple pixels~\cite{Wang:2018}. The read noise is 30 $e^-$ in the global shutter mode for Vita 5000 (dynamic range of 53 dB for the full well depth of 13700 $e^-$). We conclude that the resolution for direct detection of single X-ray photons is mainly determined by charge sharing among neighboring pixels, as confirmed by using a variable X-ray energy source (Amersham model: AMC 2084), Fig.~\ref{fig:csda1}.

Next, we consider indirect detection schemes for 10~keV and above energies, when X-rays first turn into a `cloud of visible photons' by using a scintillator. A few scintillators are summarized in Table.~\ref{Tb:list}. 
 At 20 keV, the $1/e$ X-ray attenuation lengths are 29.8, 60.9, 91.6~$\mu$m and 22.6 cm for Lu$_2$SiO$_5$(Ce) [LSO (Ce)], ZnO, (C$_6$H$_5$)$_4$PPbBr$_4$[PPh4PbBr4] and plastic C$_{10}$H$_{11}$ [EJ-228] scintillators respectively. Except for the plastic scintillator, the smaller $1/e$ attenuation length than that of silicon at $E_X = 20$ keV may allow thin-film and 2D structures ({\it esp.} for halide Perovskites)  for efficient X-ray conversion, similar to the recent work on PAL~\cite{Lee:2009}.
\begin{table}[htp]
\caption{A comparison of light yield parameters of several scintillators based on a modified Birks's model, Eq.~(\ref{BL:1}).}
\begin{center}
{\renewcommand{\arraystretch}{1.20}
\renewcommand{\tabcolsep}{0.35 cm}
\begin{tabular}{lccc}
\hline
{\bf Scintillator}  & S &$\rho$ &  ${k}_1$ \\
 & (ph/keV) & (g/cm$^{-3}$) &($\mu$m/keV) \\\hline
 LSO(Ce)  & 30 & 7.4 & 3.2$\times$ 10$^{-2}$ \\
 ZnO  & 9.0 & 5.6 & 1.5 $\times$ 10$^{-2}$ \\
PPh4PbBr4 & 6-8 &2.4& 0.01 - 0.1\\
EJ-228  & 10.2 &1.0& 0.13 \\
\hline
\end{tabular}}
\end{center}

\label{Tb:list}
\end{table}

\begin{figure}[htbp] 
  \centering
   \includegraphics[width=2.7in, angle=0]{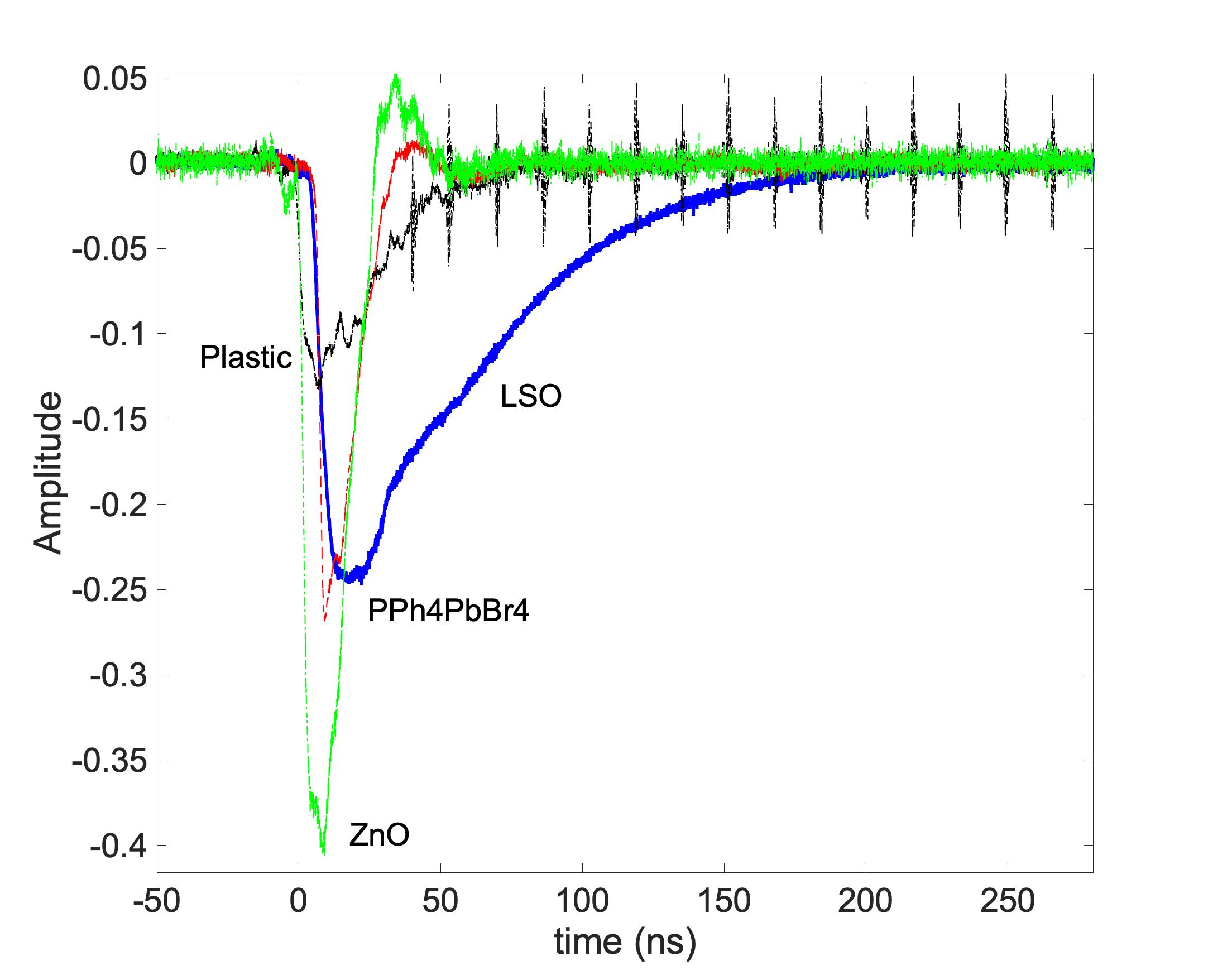} 
   \caption{Characterization of the scintillator light yield and decay time using the APS mono-energetic (29.2 keV, Sn K-edge) single-pulse X-ray in the hybrid mode. The rise time and decay time, together with the relative light yield have been obtained from the pulse shape analysis.}
  \label{fig:cp1}
\end{figure}

The relative X-ray response for four different scintillators has been measured using a Hamamatsu R2059 photomultiplier tube (Bialkali 400S photocathode, quartz window, peak QE 27\% at 390 nm) and the Argonne Advanced Photon Source (APS), Fig.~\ref{fig:cp1}. The plastic scintillator (EJ-228, 2.5 mm thick, emission peak 391 nm), ZnO (0.3 mm thick, emission peak 380 nm)~\cite{Hu:2020}, LSO (3 mm thick, emission peak 420 nm), and PPh4PbBr4 ($\sim$ 1 mm thick,  emission peak {\it est.} at 400 nm). The shape of the pulse is fitted with the function of the form $I = I_0 [\exp (- t/t_2) - \exp (- t/t_1)]$ with $t_1$ and $t_2$ being the rise and decay time respectively.

The signals from individual X-ray photons can be estimated as follows. We use the CSDA model to estimate the initial size of the photon cloud generated by photoelectrons. Fig.~\ref{fig:csda1} includes an example for LSO. The number of photons emitted is 30 ph/ keV for 1 MeV photons in LSO. The photon yield decreases by a factor $f_y< 1$ for lower energy photons. $f_y (E=30$ keV$)= 0.85$ and $f_y (E=10$ keV$)= 0.67$ in LSO~\cite{RV:1997}. At 29.2 keV X-ray photon energy, the average number of photons emitted is about 810. The critical angle is $\theta_c = asin(1/n) =0.585$ for n=1.81. The number of photons collected is about 70. For a detection efficiency of 0.3, the final number of X-ray-induced electron-hole pairs is about 20.  Transport of visible photons  from a scintillator to the sensor takes lossy steps due to refractive-index mismatching at multiple interfaces. Scintillator - CIS cover glass interface could be separated by an air gap. Additional built-in interfaces within a CIS include microlens arrays, light pipes and antireflection (AR) coating on silicon surface~\cite{TWU:2014}. Silicon has large optical refractive index ($n$) that is wavelength dependent. For example, $n$ is 5.57,  4.65, 4.30, 4.08, 3.79 at 400 nm, 452 nm, 500~nm, 550 nm, 689 nm respectively. Without AR coating, 30-40\% of the incoming light could be lost at the silicon surface alone. 

\begin{figure}[htbp] 
  \centering
   \includegraphics[width=2.7in, angle=0]{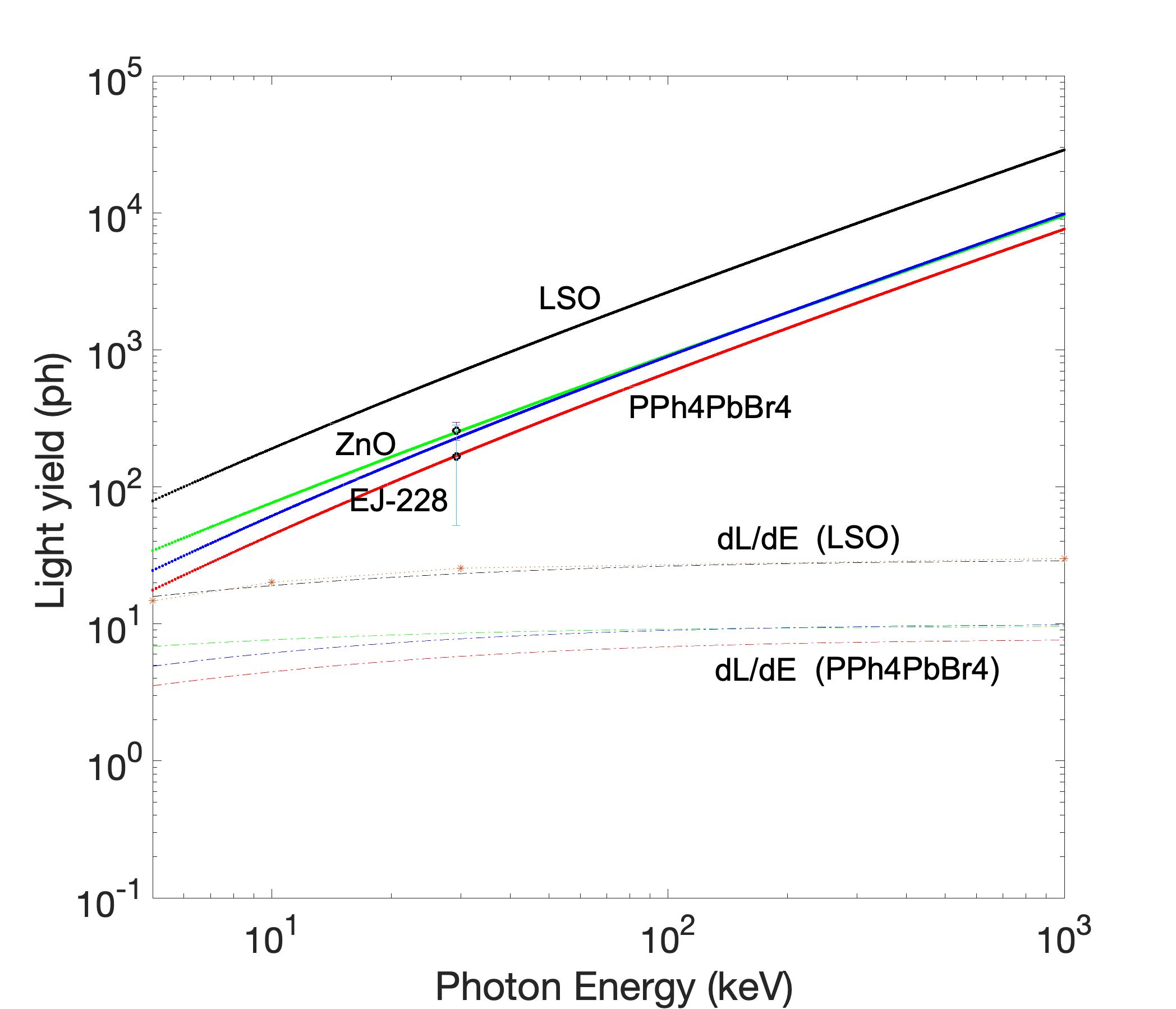} 
   \caption{Light yield model for X-ray energies from 10 to 1000~keV, when the intrinsic light yield nonproportionality is expected. The known values for LSO and EJ-228 are used to obtain the new values for ZnO and PPh4PbBr4 based on relative light intensities shown in Fig.~\ref{fig:cp1}.}
  \label{fig:Lye1}
\end{figure}

The scintillator light yield ($L_\nu$) as a function of X-ray energy uses a modified Birks's model~\cite{Birks:1964},
\begin{equation}
\frac{dL_\nu}{dE} =  \frac{S}{1 + k_1 \frac{dE}{dx}+ k_2 (\frac{dE}{dx})^2},
\label{BL:1}
\end{equation}
where $S$ is the scintillation efficiency, $dE/dx$ is the energy loss of the particle per path length, and $k_1$ is Birks's constant and material-dependent. The results here are summarized in Fig.~\ref{fig:Lye1}. The new ZnO and perovskite scintillator PPh4PbBr4 results are obtained through relative measurements shown in Fig.~\ref{fig:cp1}. The light-yield model and results will be useful in further BiPC-X optimization.

\section{Sensitivity \& resolution results} \label{sec:ex}
\begin{figure}[htbp] 
  \centering
   \includegraphics[width=3.0in,angle=0]{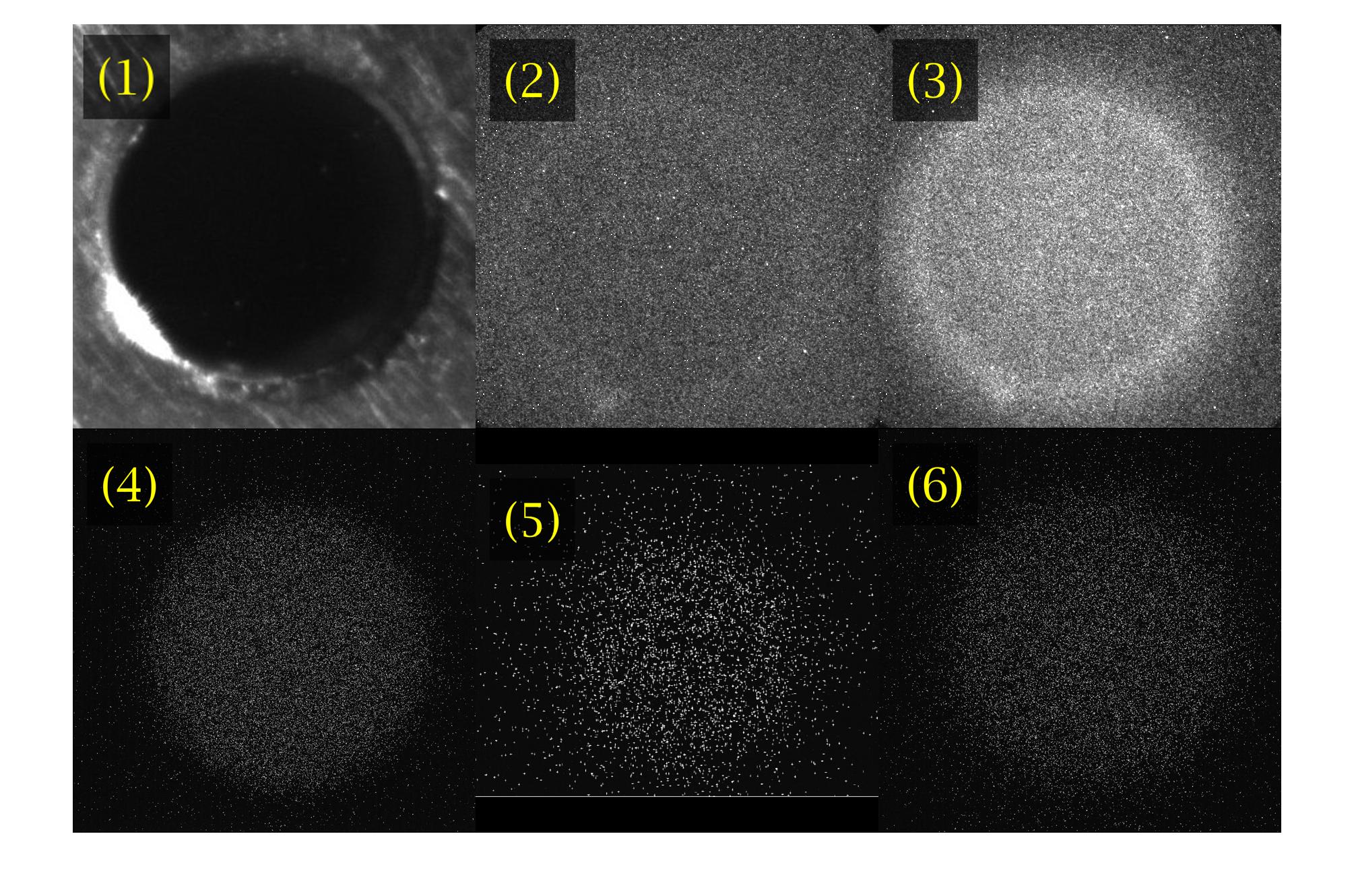} 
   \caption{(1) An Amersham variable energy X-ray source used for the single photon sensitivity test. (2) The intensified image of the X-ray source with Cu K$_\alpha$ 8.04 keV and K$_\beta$ 8.91 keV; (3) The intensified source image with Tb K$_\alpha$/K$_\beta$. (4)-(6) Direct source images from Ag, Cu and Tb K$_\alpha$/K$_\beta$ X-rays.}
  \label{fig:Amersham1}
\end{figure}

Single X-ray responses of different CIS models have been characterized using an Amersham variable energy X-ray source. Six pairs of K$_\alpha$ and K$_\beta$ lines from Cu, Rb, Mo, Ag, Ba, Tb are excited by $\alpha$ particles from $^{241}$Am radioisotope. The lowest energy is Cu K$_\alpha$ 8.04 keV. The highest energy is at Tb K$_\beta$ 50.65 keV. A few examples are shown in Fig.~\ref{fig:Amersham1}. Indirect detection results are given in panels (2) and (3) for Cu K lines and Tb K lines. An LSO in combination with various CIS did not give results with sufficiently high signal-to-noise ratio. A single-stage MCP image intensifier was able to improve the SNR as shown. Direct detection results are given in panels (4) to (6). Interactions with individual X-ray photons are clearly visible. The detection efficiency is estimated to be less than 1\% and improvements to above 10\% are planned .

\begin{figure}[thbp!] 
  \centering
   \includegraphics[width=2.5in,angle=0]{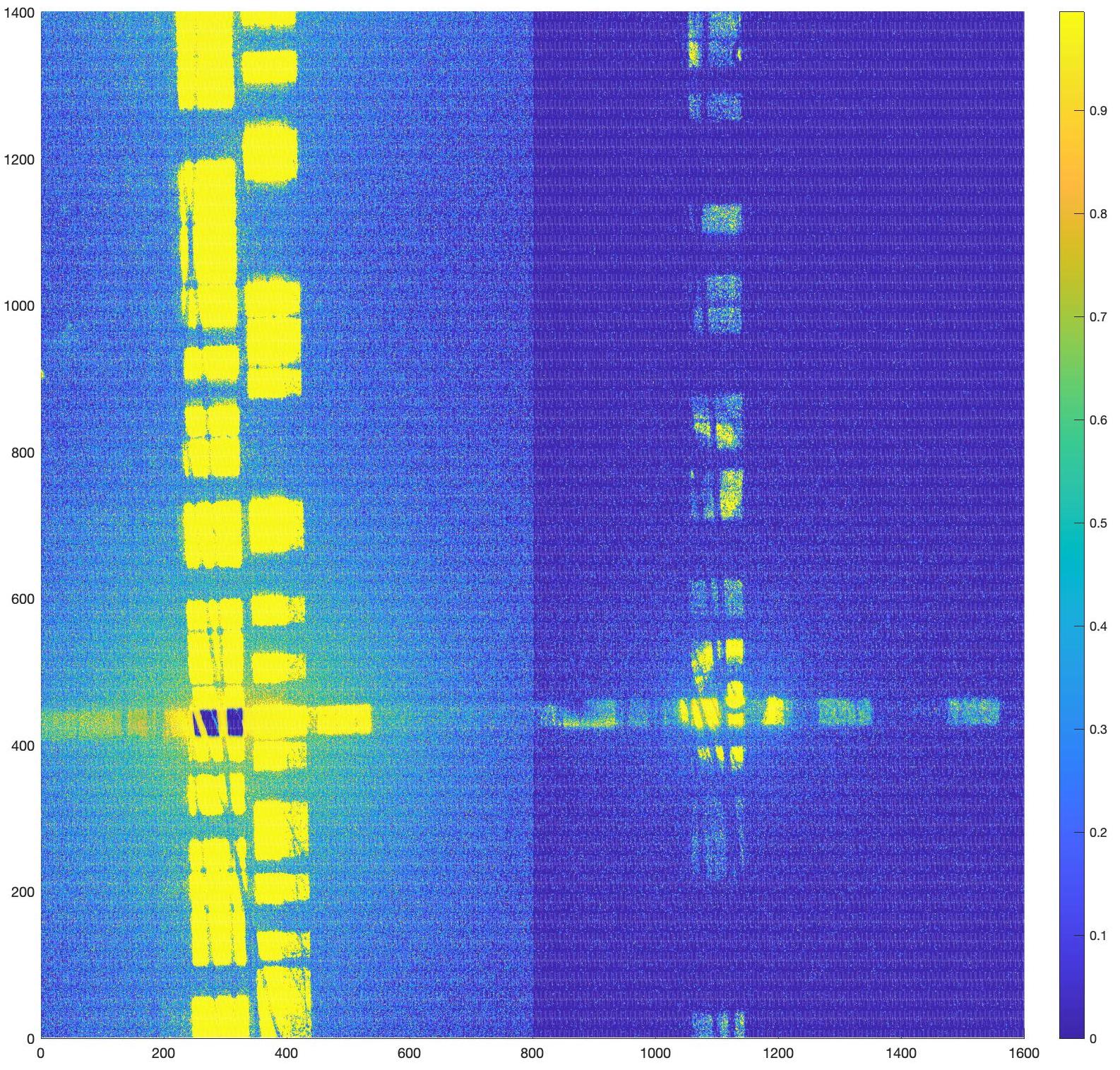} 
   \caption{X-ray images from a random wire pattern on two back-to-back stacked CIS using the APS synchrotron. The small rectangles are the spot size of the illumination.}
  \label{fig:res1}
\end{figure}

\begin{figure}[thbp!] 
  \centering
   \includegraphics[width=2.5in,angle=0]{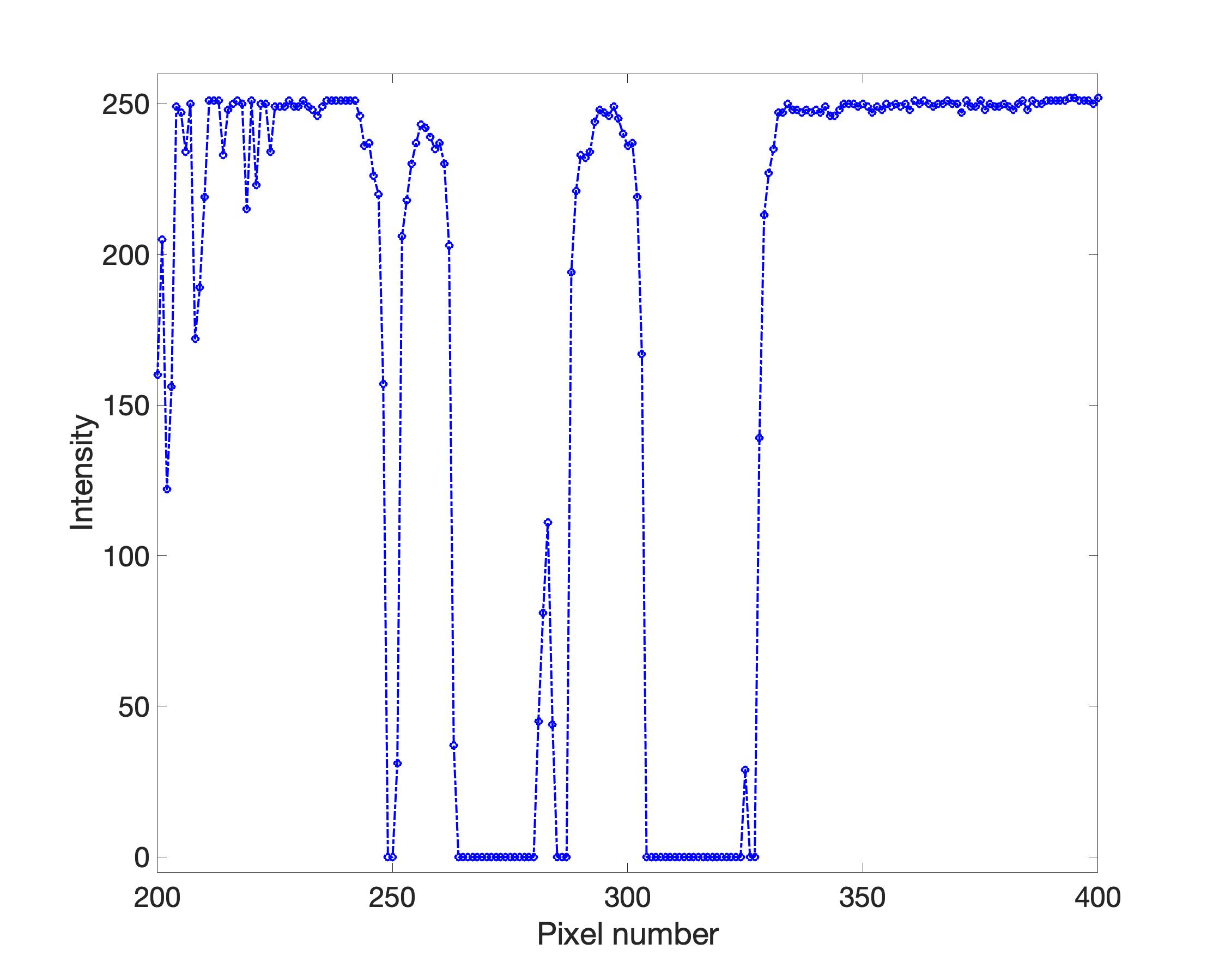} 
   \caption{Resolution test using a 20-$\mu$m diameter gold-coated tungsten wire. The FWHM for the wire projection (pixel number 280-285) is 2.7 pixels or 13 $\mu$m.}
  \label{fig:lineout1}
\end{figure}

Projection X-ray imaging using the direction detection scheme was obtained using the APS synchrotron (ID 10), Fig.~\ref{fig:res1}. Two Vita 5000 CIS were placed in a back-to-back stacked configuration along the X-ray beam path~\cite{Li:2019}. The Fresnel numbers are 2.4 $\times 10^6$ and 1.8 $\times 10^5$ (1 mm spot size) for the front and back CIS respectively. The resolution of 13 $\mu$m is obtained in Fig.~\ref{fig:lineout1} from the line-out (y = 435) measurement of Fig.~\ref{fig:res1}.


In summary, we have shown that, due to the continuing improvements in quantum efficiency, reduction in noise, and shrinking in pixel pitch, billion-pixel X-ray cameras (BiPC-X) are feasible based on commercial CMOS imaging sensors (CIS) and different tiling configurations. A 2$\times$2 planar tiling CMOS camera has been built and tested using both the laboratory X-ray sources and the APS synchrotron. BiPC-X based on direct detection is better suited for X-rays below 10 keV. Indirect detection for 10 keV and above will need CIS with single-photon sensitivity or high light yield scintillators. Further work include data handling and improvements in detection efficiency. Possible applications of BiPC-X include laser-produced and magnetically confined high-temperature plasmas when a few to 10s of keV X-rays are emitted to a wide field of view. 

We would like to thank Argonne APS ID10 staff, {\it esp.} John Katsoudas and Prof. Carlo Segre for help and coordination with scintillator measurements. The work is supported in part by the LANL Office of Experimental Sciences (C3) program (contact: Dr. Bob Reinovsky). Z. W., supported in part by the LANL/LDRD program, also wishes to thank Drs. Blas Uberuaga, Rich Sheffield, Renyuan Zhu (Caltech), Liyuan Zhang (Caltech) for stimulating discussions and help.


\end{document}